\documentclass[sigconf,natbib=true,anonymous=false]{acmart}
\usepackage{amsmath}

\usepackage{algorithm, algorithmic}

\AtBeginDocument{%
  \providecommand\BibTeX{{%
    \normalfont B\kern-0.5em{\scshape i\kern-0.25em b}\kern-0.8em\TeX}}}

\begin{document}

\title{Brain-Machine Interfaces \& Information Retrieval \\Challenges and Opportunities}

\author{Yashar Moshfeghi}
\email{yashar.moshfeghi@strath.ac.uk}
\orcid{0000-0003-4186-1088}
\affiliation{%
  \institution{NeuraSearch Laboratory\\University of Strathclyde}
  \city{Glasgow}
  \country{UK}}

\author{Niall Mcguire}
\email{niall.mcguire@strath.ac.uk}
\orcid{0009-0005-9738-047X}
\affiliation{
  \institution{NeuraSearch Laboratory\\University of Strathclyde}
  \city{Glasgow}
  \country{UK}}

\renewcommand{\shortauthors}{Moshfeghi \& Mcguire}

\begin{abstract}
The fundamental goal of Information Retrieval (IR) systems lies in their capacity to effectively satisfy human information needs - a challenge that encompasses not just the technical delivery of information, but the nuanced understanding of human cognition during information seeking. Contemporary IR platforms rely primarily on observable interaction signals, creating a fundamental gap between system capabilities and users' cognitive processes. Brain-Machine Interface (BMI) technologies now offer unprecedented potential to bridge this gap through direct measurement of previously inaccessible aspects of information-seeking behaviour. This perspective paper offers a broad examination of the IR landscape, providing a comprehensive analysis of how BMI technology could transform IR systems, drawing from advances at the intersection of both neuroscience and IR research. We present our analysis through three identified fundamental vertices: (1) understanding the neural correlates of core IR concepts to advance theoretical models of search behaviour, (2) enhancing existing IR systems through contextual integration of neurophysiological signals, and (3) developing proactive IR capabilities through direct neurophysiological measurement. For each vertex, we identify specific research opportunities and propose concrete directions for developing BMI-enhanced IR systems. We conclude by examining critical technical and ethical challenges in implementing these advances, providing a structured roadmap for future research at the intersection of neuroscience and IR.
\end{abstract}



\begin{teaserfigure}
  \includegraphics[width=\textwidth, height=5cm]{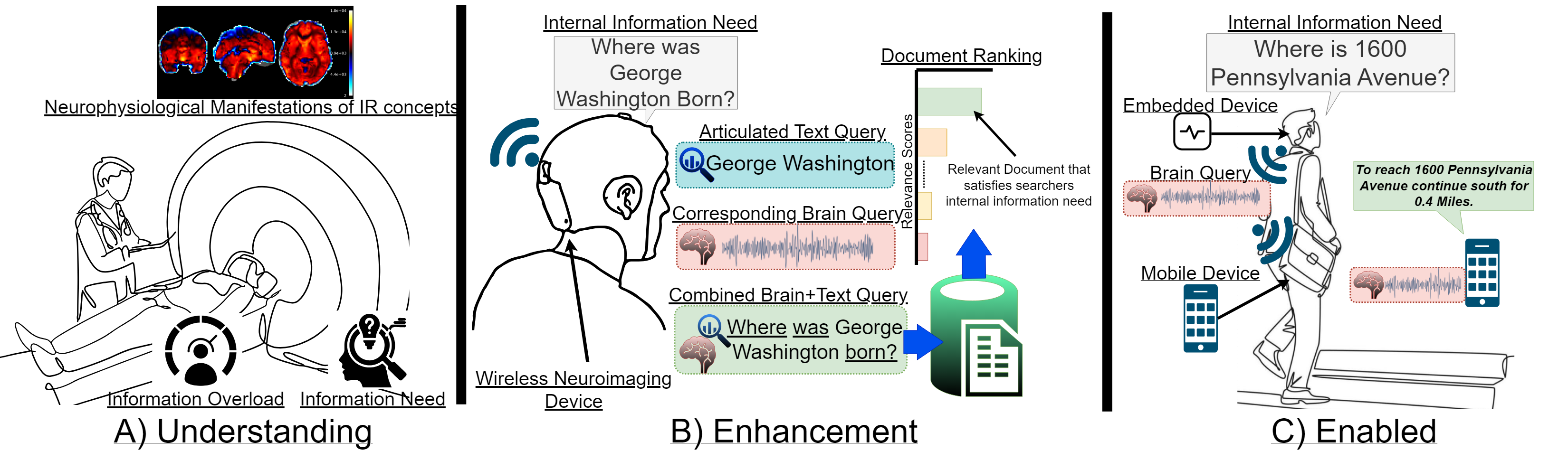}
  \caption{BMI Opportunities in Information Retrieval: (A) Understanding neural correlates of IR concepts through BMI, (B) Neurophysiological Enhancement of IR, and (C) Neurophysiological Enabled IR}
  \Description{BMI Opportunities in Information Retrieval: (A) Understanding neural correlates of IR concepts through BMI, (B) Neurophysiological Enhancement of IR, and (C) Neurophysiological Enabled IR}
  \label{fig:teaser}
\end{teaserfigure}

\maketitle

\section{Introduction}
\label{Introduction}

\textit{What if IR systems could understand and satisfy our information needs before we even articulate them?} Despite significant advances in search technology, including Conversational Search and Generative AI assistants \cite{deldjoo2021towards, mo2024survey}, IR systems remain fundamentally reactive, dependent on explicit user actions to infer intent and relevance, e.g. queries, clicks, dwell time, and browsing patterns \cite{agichtein2006improving, white2007investigating}, creating a fundamental gap between system capabilities and users' cognitive processes. This gap is a result of the inherent complex cognitive states that shape information seeking, i.e. the moment an information need (IN) crystallises in a user's mind, the instantaneous judgement of a document's relevance, or the subtle shifts in understanding as users process search results \cite{belkin1982ask, kuhlthau2005information}. This is further pronounced in scenarios with ill-defined information needs such as complex or exploratory search  \cite{marchionini2006exploratory}, where users may struggle to verbalise their needs effectively \cite{belkin1982ask, kuhlthau2005information}. As such, while these technological advancements have improved interactive and context-aware search methods \cite{white2013enhancing}, they still require expressed intent, failing to capture the rich cognitive dimensions of information seeking that occur before any observable behaviour. Moreover, traditional IR interfaces remain inaccessible to users with motor disabilities \cite{pasqualotto2015usability}, cognitive impairments \cite{lazarou2018eeg}, or limited literacy, as they depend on explicit text or voice interactions, reinforcing usability barriers.

At the core of IR is the challenge of understanding and satisfying users' information needs \cite{belkin1982ask, ingwersen1996cognitive, kuhlthau2004seeking}. To truly achieve this, IR systems need to go beyond processing explicit queries and implicit behavioural signals to interpreting the underlying cognitive states that drive information seeking. While IR systems have evolved substantially over decades \cite{baeza1999modern, white2013enhancing}, they remain fundamentally limited by their inability to directly measure and respond to users' internal cognitive states \cite{ingwersen1996cognitive, moshfeghi2016understanding}. This underscores the need for cognition-aware IR systems that can directly interpret cognitive intent, making information access more equitable, efficient, and inclusive. We now stand at a transformative juncture: Brain-Machine Interface (BMI) technologies now offer unprecedented potential to directly measure and interpret these previously inaccessible aspects of information-seeking behaviour \cite{nicolas2012brain, moshfeghi2016understanding, moshfeghi2021neurasearch, mostafa2016deepening}. 

Research in neuroscience and computational neuroscience has demonstrated that search behaviour is directly linked to measurable brain activity, including query formulation, relevance assessment, and cognitive load, as detected through EEG, fMRI, and MEG \cite{jacucci2019integrating, moshfeghi2016understanding, gwizdka2017temporal, moshfeghi2013understanding, mcguire2024prediction, 10.1145/3477495.3532082, 10.1145/3485447.3511966}. Studies confirm that relevance decisions can be predicted up to 500 milliseconds before users consciously express them \cite{eugster2014predicting}, and cognitive load variations can be continuously tracked \cite{gwizdka2017temporal}, offering real-time insights into user engagement and retrieval difficulty. These findings suggest that BMI-enhanced retrieval models offer opportunities to refine search results, predict frustration, and even anticipate emerging information needs before users explicitly formulate them \cite{moshfeghi2019towards}.

Empirical research further validates the feasibility of BMI enhanced IR systems. Recent advances in non-invasive EEG-based neurophysiological sensing now enable real-time cognitive state monitoring \cite{mihajlovic2015wearable}, which can be integrated into retrieval models to detect frustration, predict relevance, and dynamically refine search queries \cite{eugster2016natural, eugster2014predicting, ye2023relevance, davis2021collaborative}. This development fundamentally transforms IR from passive document retrieval to cognition-aware search, where systems respond not just to explicit user actions but to real-time cognitive feedback. Instead of requiring users to struggle with query formulation, BMI-enhanced search engines could sense uncertainty, knowledge gaps, and engagement levels, dynamically adjusting retrieval strategies before users take action \cite{moshfeghi2016understanding, ye2024query, del2021dealing}. The convergence of BMI technology with IR represents more than just a technical advancement. It offers a fundamental re-conceptualisation of how we understand and support human information-seeking behaviour \cite{wilson2000human, mostafa2016deepening}. 

In this paper, we examine opportunities and challenges for integrating BMI capabilities into IR research, organizing our analysis around three key vertices: (1) understanding the neural correlates of information-seeking behaviour to advance theoretical models of search behaviour, (2) enhancing existing IR systems through contextual integration of brain signals, and (3) developing proactive capabilities that can detect and respond to information needs at their inception \cite{sitaram2017closed}. For each vertex, we identify specific research opportunities and propose concrete directions for developing BMI-enhanced IR systems.

Whilst BMI-enhanced IR has great potential, several significant challenges must be addressed. Brain signal processing remains a major hurdle, as real-world brain signal acquisition suffers from high noise levels and requires advanced filtering techniques \cite{lotte2018review, kingphai2021eeg}. Additionally, BMI-enhanced IR must account for substantial variability in brain responses across individuals \cite{schirrmeister2017deep, 10.1145/3483382.3483387}, making it difficult to generalise models without personalised adaptations. Ethical and privacy concerns also present serious obstacles, as brain data is highly personal, raising issues about consent, security, and responsible application of neurotechnologies in IR systems \cite{sitaram2017closed, gwizdka2013applications}.

Advancing BMI-enhanced IR requires collaboration across multiple disciplines, including IR, neuroscience, machine learning, and human-computer interaction, to ensure that cognition-aware search is both effective and ethically responsible \cite{gwizdka2019introduction, moshfeghi2021neurasearch}. The integration of BMI with search represents a significant shift in human-information interaction \cite{marchionini2008human}, where search engines evolve from passive tools into cognition-aware assistants capable of understanding and responding to intent at the level of thought itself.

The paper is organized as follows. Section \ref{Intro-To-Neurotech} introduces BMI principles for IR, followed by Section \ref{BMI&IR}, which examines the theoretical foundations for brain measurement in IR contexts. Sections \ref{Contextual} and \ref{NeuroIR} present our identified opportunities for BMI integration in existing IR systems. Section \ref{challenges} discusses technical challenges and ethical considerations, while Section \ref{Summary} concludes with recommendations for advancing this research agenda.

\section{Introduction to Neurotechnology}
\label{Intro-To-Neurotech}
This section provides an overview of key neuroimaging approaches and their applications in Brain-Machine Interfaces (BMIs), establishing the technical foundation for their integration with IR systems.

{\bf Neuroimaging Technologies.} Neuroimaging technologies can be broadly categorised into invasive and non-invasive approaches, each offering distinct capabilities for measuring brain activity \cite{gaillard2017invasive, kim2021brief, hill2012recording}. Invasive methods, which require surgical implantation of recording devices, provide the highest spatial and temporal resolution for brain signal acquisition. These approaches typically utilise microelectrode arrays implanted directly into the brain tissue or electrocorticography (ECoG) arrays placed on the cortical surface \cite{lebedev2017brain, hill2012recording}. While these methods achieve exceptional precision in measuring individual neuron activity, their clinical requirements present a substantial barrier of entry in comparison to their non-invasive counterparts, limiting their applicability primarily to medical/lab-based research contexts \cite{hochberg2012reach}. Non-invasive neuroimaging techniques, which measure brain activity without requiring surgical intervention, offer more practical approaches for studying cognitive processes in information-seeking contexts. Functional Magnetic Resonance Imaging (fMRI) \cite{moshfeghi2016understanding, smith2004overview} is a non-invasive method and utilises powerful magnetic fields to detect changes in blood oxygenation levels associated with brain activity \cite{sitaram2007fmri}. This technique provides millimetre-scale spatial resolution throughout the entire brain, enabling precise mapping of brain activity patterns during complex cognitive tasks. However, fMRI's temporal resolution is limited by the inherent delay in blood flow changes, and its infrastructure requirements—including large, stationary scanning equipment—restrict its use primarily to controlled laboratory settings \cite{moshfeghi2016understanding, mcguire2024prediction}. Magnetoencephalography (MEG) offers an alternative approach, measuring the magnetic fields produced by brain activity using highly sensitive sensors positioned around the head \cite{da2013eeg}. MEG combines good spatial resolution with excellent temporal precision, capable of detecting brain signals at the millisecond scale. This temporal accuracy makes it particularly valuable for studying the rapid cognitive processes involved in information processing and decision making. However, like fMRI, MEG requires sophisticated infrastructure and carefully controlled environments, limiting its practical applications \cite{wikswo1993future}. Among non-invasive technologies, Electroencephalography (EEG) \cite{da2013eeg} has emerged as a particularly promising approach for IR applications \cite{mcguire2024prediction, michalkova2024understanding, zhang2024eeg}, offering a balance of practical utility and measurement capability. EEG systems measure electrical potentials generated by brain activity through electrodes placed on the scalp \cite{mihajlovic2015wearable}. While their spatial resolution is limited compared to other methods, modern EEG systems offer millisecond-scale temporal resolution and have benefited from significant advances in dry electrode technology \cite{lopez2014dry} and wireless recording capabilities \cite{niso2023wireless}. These developments have made EEG increasingly practical for studying cognitive processes in more naturalistic settings.

{\bf Brain-Machine Interfaces (BMIs).} BMIs enable direct communication between the human brain and external devices through the measurement and interpretation of brain activity \cite{wolpaw2002brain}. These systems function by converting brain activity patterns into commands that can control computers, devices, or communication systems. BMIs operate through three main stages: signal acquisition using neuroimaging methods, signal processing to extract relevant features, and translation of these signals into specific commands or outputs \cite{ramadan2017brain}. Recent advances in machine learning have significantly enhanced BMI capabilities. Modern systems use sophisticated architectures, including transformer models for temporal patterns \cite{kostas2021bendr} and convolutional networks for spatial features \cite{schirrmeister2017deep}. Many of these approaches share similarities with methods already established in IR, such as sequence modelling for query understanding and representation learning for document encoding. BMI systems employ various learning strategies, including supervised learning for direct signal-to-command mapping, reinforcement learning for adaptive control, and self-supervised learning for robust feature extraction from unlabelled brain data \cite{roy2019deep}. These advances have enabled BMIs to detect and interpret complex cognitive states with increasing accuracy, moving beyond basic motor control to applications in emotion recognition and intent prediction. For IR systems, BMIs offer the potential to directly measure users' cognitive states during search tasks. Studies have shown that BMIs can detect relevance judgments \cite{eugster2016natural}, measure cognitive load \cite{gwizdka2017temporal}, and identify emerging information needs \cite{moshfeghi2016understanding}. While invasive BMIs provide the highest signal quality, non-invasive approaches, particularly EEG-based systems, present more practical solutions for IR applications \cite{jacucci2019integrating}. However, significant challenges remain in translating these capabilities to practical applications, including managing signal variability and achieving real-time processing speeds \cite{lotte2018review}. Despite the demonstrated success of BMI methods across various domains, their potential to enhance IR systems remains largely unexplored. 
In the following section, we examine opportunities that these brain measurement capabilities could bring to IR. 

\section{Opportunities for BMI in IR}
\label{BMI&IR}
The integration of BMIs with IR systems represents a transformative opportunity to bridge the gap between users' cognitive processes and information access systems. While traditional IR relies on explicit queries and observable behaviours, BMI integration offers unprecedented insight into the brain mechanisms underlying information seeking. We identify opportunities for integration oriented around three fundamental vertices: (1) understanding neural correlates of core IR concepts such as IN realisation, relevance assessment, and satisfaction judgment; (2) enhancing existing IR systems through real-time cognitive feedback in both search and recommender systems; and (3) developing neuro-proactive capabilities that can detect and respond to information needs at their inception. Throughout this section and the following sections, we propose specific Research Directions (RDs) that highlight promising avenues for future investigation. These Research Directions represent concrete paths forward for the IR community to explore the integration of BMI technologies into information retrieval systems. Our analysis emphasizes practical opportunities achievable with current BMI technology, supported by evidence from both laboratory studies and emerging applications.

\subsection{Understanding Neural Correlates of IR}
\label{Understanding}
Traditional IR has relied on user studies, think-aloud protocols, and interaction logs to understand how users satisfy their information needs in an information-seeking process \cite{kelly2009methods}. While foundational, these methods offer limited insight into real-time cognitive processes underlying such processes. Advances in neuroimaging techniques (e.g. fMRI, EEG, MEG) now enable direct brain activity measurement during search tasks \cite{moshfeghi2013understanding, pinkosova2023moderating, pinkosova2020cortical, moshfeghi2016understanding, michalkova2022information}. This interdisciplinary approach bridges neuroscience and IR by capturing brain activity during information seeking, revealing how INs emerge, evolve, and drive search behaviour \cite{moshfeghi2021neurasearch}. Integrating neurophysiological data with traditional methods allows researchers to refine IR models and develop cognition-aware search systems. 

\subsubsection{Information Needs} 
Understanding how information needs (INs) emerge and evolve is key to advancing cognition-aware IR. The Neuropsychological Model of IN Realisation \cite{moshfeghi2019towards} identifies three core processes: (1) memory retrieval, assessing knowledge availability; (2) information flow regulation, enabling cognitive transitions; and (3) high-level perception, detecting knowledge gaps. The posterior cingulate cortex orchestrates these functions, with heightened activity signalling IN awareness \cite{moshfeghi2016understanding, moshfeghi2019towards, mcguire2024prediction}. EEG studies confirm that IN realisation precedes conscious awareness, with specific brain activity patterns occurring at different time points: early electrical responses (known as the N1-P2 complex, observable 100-200 milliseconds after stimulus presentation) marking the initial awareness processes, while later brain activity patterns (including the N400 and P600 components, occurring 400-600 milliseconds after stimulus) reflect memory control mechanisms that help distinguish between known and unknown information states \cite{michalkova2022information}. Michalkova et al. (2022) found that EEG patterns also differentiate correct from incorrect recall, providing a foundation for cognition-aware IR models that anticipate user needs before explicit search behaviour begins \cite{michalkova2022information}. The feasibility of real-time cognition-aware IR has been demonstrated by McGuire and Moshfeghi \cite{mcguire2024prediction}, who showed EEG-based models predict IN realisation with up to 90.1\% accuracy. These findings, combined with fMRI evidence of pre-query brain activity linked to emerging IN states \cite{moshfeghi2016understanding}, support proactive IR systems that anticipate cognitive states before search initiation.

Neuroimaging research reveals distinct cognitive shifts across search stages. fMRI studies \cite{moshfeghi2016understanding, moshfeghi2013understanding, moshfeghi2019towards} show varying brain activation patterns from IN to Satisfaction Judgment, reflecting cognitive and emotional processing. \citet{ji2024characterizing} integrate EEG, electrodermal activity (EDA), and pupillary responses, confirming that IN is marked by cognitive load, Query Formulation requires attentional control, and Relevance Judgment elicits emotional responses. These findings suggest IR systems should dynamically adapt to users' evolving cognitive states. Neuroscientific insights mark a notable shift beyond query-based IR. Future systems could detect knowledge gaps, assess memory recall, and predict search initiation based on cognitive load and uncertainty, enabling intelligent, adaptive search experiences.

\textbf{RD1:} \textit{Expand the theoretical framework of IN realisation by mapping its neural correlates across diverse cognitive states, incorporating multimodal neuroimaging (EEG, fMRI, MEG) to capture transitions between knowledge awareness, uncertainty, and search intent formation in real-world search scenarios.}

\subsubsection{Relevance} 
Understanding how users perceive and assess relevance is fundamental to advancing cognition-aware IR. fMRI studies reveal that relevance processing engages the inferior parietal lobe, inferior temporal gyrus, and superior frontal gyrus—brain regions linked to attentional control, semantic integration, and decision-making \cite{moshfeghi2013understanding, moshfeghi2016understanding}. Research supporting Saracevic's multidimensional relevance framework demonstrates distinct pathways for topical and situational relevance \cite{saracevic1996relevance, moshfeghi2013cognition, allegretti2015relevance}. EEG studies further validate this by identifying specific brain responses (known as event-related potentials or ERPs) that occur at different time points after viewing content. These include attention-related responses (around 300 milliseconds after stimulus), semantic processing responses (around 400 milliseconds), and contextual integration responses (around 600 milliseconds), all serving as neural markers of relevance perception, indicating that assessments occur before conscious articulation \cite{pinkosova2023moderating}. EEG research has also uncovered the temporal dynamics of relevance assessment, with brain responses peaking within 500–800ms post-stimulus \cite{eugster2014predicting}. Early visual processing (P100) plays a role in initial relevance detection \cite{allegretti2015relevance}, while \citet{kim2018does} demonstrated that ERP components related to semantic congruency (N400) and memory integration (P600) distinguish between relevant and non-relevant documents. 

These findings challenge traditional IR models that assume users consciously assess relevance only after explicit document review, instead suggesting that subconscious cognitive processing shapes relevance judgments before users verbalise them. However, challenges remain. Neurophysiological response variability across individuals necessitates more generalisable brain-driven relevance models. While EEG offers high temporal resolution, its spatial precision is limited, requiring multimodal neuroimaging approaches to fully capture relevance processing \cite{moshfeghi2013understanding, pinkosova2023moderating}. While technical progress is promising, challenges remain in making these systems robust enough for practical applications. Neuroscientific insights into relevance processing open avenues toward cognition-aware IR, where systems anticipate relevance judgments, refine rankings, and dynamically present information based on users' cognitive states. These insights show promise to make search experiences more intuitive, adaptive, and personalised \cite{eugster2016natural, pinkosova2022revisiting, allegretti2015relevance, ye2024relevance}.

\textbf{RD2:} \textit{Develop cognition-aware relevance models that integrate real-time brain signals (EEG-based ERPs, fMRI activations) to uncover how relevance is perceived in dynamic, real-world search contexts, accounting for evolving task complexity, time pressure, and individual cognitive variability.}


\subsubsection{Satisfaction and Search Termination}
\label{satisfaction}
Satisfaction is a crucial factor in IR \cite{siro2022understanding, 10.1145/3503161.3548258}, influencing search termination, user engagement, and perceived system effectiveness \cite{wang2014modeling}. Neuroscientific research now offers direct insights into how users determine when they have fulfilled an IN \cite{paisalnan2021towards}. fMRI studies reveal that satisfaction engages brain regions linked to decision-making and emotional evaluation, including the anterior cingulate, superior frontal gyrus, insula, and inferior frontal gyrus \cite{paisalnan2021towards}. These regions differentiate satisfaction from ongoing search states and support the development of cognition-aware IR systems that adapt dynamically to users' evolving cognitive states \cite{paisalnan2022neural}. Recent fMRI investigations have mapped the brain transitions from IN realisation to relevance judgment and satisfaction judgment in a search process, showing distinct cognitive pathways for each phase. 

While IN activates attention, working memory, and planning regions, relevance judgment and satisfaction judgment engage semantic processing and evaluative networks \cite{paisalnan2022neural}. These insights refine information foraging models, conceptualising search as an evidence accumulation process where users continue searching until a cognitive threshold is reached, at which point brain activity in decision-making regions declines, and satisfaction emerges \cite{paisalnan2021towards}. This cognitive threshold model shares interesting parallels with Herbert Simon's concept of ``satisficing,'' where decision-makers accept solutions that meet a threshold of adequacy rather than pursuing optimal solutions \cite{simon1956rational}. While the relationship between satisfaction judgment in search and satisficing behaviour remains under-explored, neurophysiological methods offer promising approaches to investigate whether the cognitive mechanisms underlying these processes share common neural substrates. Understanding this relationship could significantly advance our theoretical models of search termination behaviour and enhance the development of systems that better accommodate users' natural decision-making processes.

Distinguishing satisfaction from frustration is vital for enhancing IR experiences. Satisfaction correlates with reduced cognitive load and heightened reward-related activity, while frustration manifests as increased amygdala and prefrontal cortex activation, signalling cognitive dissonance \cite{paisalnan2022neural}. Recognising these distinctions may allow cognition-aware IR systems to anticipate user frustration and intervene pre-emptively by adjusting retrieval strategies based on inferred cognitive states.

\textbf{RD3:} \textit{Advance the understanding of satisfaction realisation by examining its neurophysiological underpinnings across diverse task types, user expertise levels, and external influences, developing cognition-aware IR models that predict and adapt to user satisfaction in real-world, high-stakes search scenarios.}

\subsubsection{Contextual Human Factors}
\label{human-factors}
Beyond satisfaction, neurophysiological research has revealed how cognitive and emotional states fundamentally shape information-seeking behaviour. Studies demonstrate that different search tasks create distinct cognitive load patterns \cite{gwizdka2010distribution, crescenzi2014too}, with factors such as task complexity, time pressure, and user expertise significantly influencing brain responses. This cognitive variability manifests not only in task execution but also in how users process and interact with search interfaces \cite{cutrell2007you}, suggesting the need for systems that can dynamically adapt to users' cognitive states \cite{white2013enhancing}. The emotional dimensions of search behaviour are equally critical. Neuroscientific research has identified distinct brain signatures for various emotional states during search, including frustration, engagement, and uncertainty \cite{moshfeghi2013cognition, alarcao2017emotions}. Importantly, these brain indicators often precede observable behaviours, offering opportunities for proactive system intervention \cite{hassan2014struggling}. For instance, detecting early signs of user frustration could enable systems to adjust result presentation or provide additional support before users abandon their search \cite{white2009characterizing}.

The integration of these cognitive and emotional insights into IR systems represents a significant advancement beyond traditional behavioural approaches \cite{jacucci2019integrating}. While current research demonstrates the feasibility of measuring neural correlates during search tasks \cite{moshfeghi2016understanding, eugster2014predicting}, the greater potential lies in developing systems that actively respond to users' cognitive and emotional states \cite{gwizdka2017temporal}. This approach requires rethinking traditional evaluation methods \cite{kelly2009methods} to incorporate metrics that capture cognitive load, emotional engagement, and search effectiveness \cite{sanderson2010test}. Future research should prioritise understanding how different cognitive states influence search performance \cite{eugster2016natural}, developing more sophisticated models of user engagement, and creating adaptive systems that provide personalized support based on real-time neurophysiological feedback \cite{white2013enhancing}

\textbf{RD4:} \textit{Investigate how different cognitive and emotional states (including frustration, engagement, and uncertainty) influence search behavior and performance, developing models that can predict and respond to these states to create more personalized search experiences.}

\subsection{Neurophysiological Enhancement of IR}
\label{Contextual}
While BMIs present transformative potential for better understanding physical manifestations of IR concepts, immediate opportunities exist for enhancing current IR systems through targeted integration of neurophysiological data. This section examines how brain signals can address specific limitations in two fundamental components of modern information access: \textit{search} and \textit{recommendation}. Search system enhancements focus on bridging the gap between users' internal information needs and their external expressions, potentially transforming how users interact with search interfaces. Recommender system applications leverage neural signals to create more accurate and responsive personalization models, addressing critical challenges such as cold-start problems and real-time preference detection. We emphasize practical implementations within existing IR frameworks while acknowledging the technical challenges and research opportunities that emerge from integrating brain signals into operational systems. 

\subsubsection{Enhancement of Search}
\label{search-enhancement}
Search systems face fundamental limitations in their ability to effectively capture and respond to users' information needs \cite{belkin1982ask, ingwersen1996cognitive}, despite significant advances in search technology \cite{white2013enhancing, mo2024survey}. These limitations manifest primarily in three critical areas: query formulation and refinement \cite{taylor1968question, belkin1982ask}, search task adaptation \cite{white2009exploratory, marchionini2006exploratory}, and results presentation \cite{agichtein2006improving, joachims2017accurately}. Traditional search interfaces rely heavily on explicit textual input, creating a significant cognitive burden as users struggle to externalize their information needs into effective queries \cite{belkin1982ask, taylor1968question}. This challenge is particularly acute when information needs are ill-defined or evolving, leading to a persistent semantic gap between users' mental models and their textual expressions \cite{ingwersen1996cognitive, moshfeghi2016understanding}.

Query formulation and refinement present significant opportunities for neurophysiological enhancement of IR systems through two primary pathways: enhancement of existing query mechanisms and direct brain querying \cite{belkin1982ask, jacucci2019integrating, eugster2016natural}. Traditional IR approaches rely heavily on query representations, whether sparse (e.g., boolean, BM25) or dense (neural embeddings) \cite{robertson2009probabilistic, lin2021pretrainedtransformerstextranking, Nogueira2019PassageRW, baeza1999modern, agichtein2006improving}. While both paradigms have proven effective, they fundamentally depend on explicit textual input, introducing an inherent semantic gap between users' cognitive models and their expressed queries \cite{belkin1982ask, taylor1968question, ingwersen1996cognitive, kuhlthau2004seeking}. Classic query expansion methods, from relevance feedback \cite{rocchio1971relevance, agichtein2006improving} to pseudo-relevance feedback \cite{lavrenko2017relevance, croft2001relevance}, attempt to bridge this gap by leveraging document collections or user interactions \cite{white2013enhancing, teevan2010potential}. Modern neural approaches to query expansion \cite{nogueira2019document} have further advanced this capability through contextual understanding and semantic matching. However, these methods remain constrained by their reliance on initial query terms, which often inadequately capture the richness of users' internal information needs \cite{belkin1982ask}.
Recent advances in brain decoding suggest promising solutions through the integration of brain-derived semantic information \cite{mitchell2008predicting, huth2016natural, Yang2024MADMM, Yang2024NeuSpeechDN, 10.1145/3488560.3502185} into query representations \cite{ye2024query, ye2023language, eugster2014predicting}. Contemporary brain decoding techniques have been shown to extract complex semantic features from brain activity \cite{mitchell2008predicting, huth2016natural, Yang2024MADMM, Yang2024NeuSpeechDN}, potentially enabling systems to capture aspects of information needs that users struggle to verbalize \cite{ye2024query}. This capability could enhance both sparse and dense query representations in several ways. For sparse representations, brain signals could identify relevant terms not explicitly mentioned in the user's query but present in their cognitive processing. In dense representations, brain-derived semantic features could be integrated directly into the embedding space, potentially improving the alignment between query vectors and users' intended search targets \cite{ye2024query}.

A more transformative approach involves bypassing explicit query formulation entirely through direct brain querying. As discussed, advanced neurophysiological decoding techniques have demonstrated increasingly robust capabilities in extracting semantic information from brain signals \cite{d2024decoding, sato2024scaling}. These advances suggest the possibility of implementing continuous pre-engagement monitoring, where systems actively process brain signals during normal task execution. Through established semantic decoding methods \cite{ye2023language,mitchell2008predicting, huth2016natural, Yang2024MADMM, Yang2024NeuSpeechDN}, these signals could be translated into dynamic query representations that more accurately reflect users' internal states.
The technical implementation of BMI-enhanced querying presents several architectural possibilities. One approach involves mapping decoded brain patterns to existing semantic spaces used in modern retrieval models. This could involve training cross-modal encoders that align brain activity patterns with textual semantic representations \cite{huth2016natural}. Alternatively, retrieval models could be adapted to operate directly on brain signal patterns, potentially enabling more direct matching between cognitive states and document representations \cite{eugster2014predicting}. Recent work demonstrates the feasibility of training models to predict relevance judgments from brain signals \cite{eugster2016natural}, suggesting similar approaches could be applied to query formulation.

The integration of brain signals into query processing offers several distinct technical advantages. First, it enables more equitable information access by providing alternative interaction pathways for users with motor impairments or those who struggle with traditional text input \cite{pasqualotto2015usability, lazarou2018eeg}. Second, it could enable a more precise capture of search intent, particularly valuable for complex or exploratory search tasks where users struggle with query articulation \cite{marchionini2006exploratory, white2009exploratory}. Third, it could facilitate continuous query refinement based on unconscious relevance judgments, and shifting information needs \cite{moshfeghi2016understanding, eugster2014predicting}. Fourth, it may allow automatic task adaptation through brain pattern classification, enabling systems to dynamically adjust retrieval strategies based on detected cognitive states \cite{gwizdka2017temporal, jacucci2019integrating}. These capabilities are particularly relevant for scenarios involving complex information needs, where traditional query formulation often falls short \cite{belkin1982ask, li2008faceted}.


\textbf{RD5:} \textit{Develop methods for effectively integrating neurophysiological signals into existing search engines, focusing on query enhancement, result re-ranking, and interface adaptation without requiring fundamental changes to system architecture.}

\subsubsection{Enhancement of Recommender Systems}
\label{recommender-enhancement}
Recommender systems represent a critical component of modern information access~\cite{Schafer2007CollaborativeFR}, yet face persistent challenges in accurately modelling user preferences and providing timely, contextually relevant recommendations~\cite{herlocker2004evaluating, karatzoglou2010multiverse}. Traditional approaches rely on historical interaction data and explicit user actions~\cite{bennett2012modeling}, creating fundamental limitations in their ability to capture emerging interests and immediate information needs~\cite{white2013enhancing, teevan2010potential}. The integration of neurophysiological data offers promising opportunities to address these limitations through direct measurement of users' cognitive responses to recommendations~\cite{eugster2016natural, davis2021collaborative, zhang2024eeg} and real-time preference detection~\cite{jacucci2019integrating, zhang2024eeg}.

Brain signals offer significant potential for advancing personalisation in recommender systems by addressing several fundamental limitations of current approaches~\cite{jacucci2019integrating, eugster2016natural}. Contemporary personalisation systems face three critical challenges where BMI integration could provide transformative solutions. First, the inherent latency between a user's cognitive response and their observable actions creates a significant temporal gap in adaptation mechanisms~\cite{saracevic1997stratified, gwizdka2019introduction}. This delay means that by the time traditional systems detect preference shifts through behavioural signals, the user's actual interests may have already evolved~\cite{white2013enhancing}. Second, behavioural indicators such as clicks and dwell time suffer from fundamental ambiguity - they may reflect genuine interest, confusion, or various forms of interaction bias, making it difficult to draw reliable conclusions about user preferences~\cite{joachims2017accurately, yilmaz2014relevance}. Third, and perhaps most significantly, existing approaches lack the capability to measure the underlying cognitive context that shapes how users evaluate and process recommendations~\cite{kelly2009methods, white2009exploratory}, limiting their ability to provide truly context-aware personalisation~\cite{teevan2010potential}.

Recent neuroscientific research has demonstrated robust approaches for addressing personalisation limitations through direct measurement of cognitive states during information interaction~\cite{moshfeghi2016understanding, eugster2016natural}. Neuroimaging studies, particularly using fMRI, have identified specific activation patterns in the anterior cingulate cortex and prefrontal regions that correlate with preference formation and decision-making processes~\cite{moshfeghi2013understanding, moshfeghi2013cognition}. These brain markers emerge rapidly, with studies showing reliable detection within 200-300ms post-stimulus~\cite{eugster2014predicting, pinkosova2020cortical}, significantly preceding observable behavioural responses. Complementary EEG research has further validated the feasibility of detecting preference-related signals during initial information exposure~\cite{allegretti2015relevance, pinkosova2023moderating}, establishing a foundation for real-time preference modelling.
A particularly promising application of these neurophysiological insights lies in addressing the cold-start problem, a persistent challenge in recommender systems~\cite{white2013enhancing, herlocker2004evaluating}. Traditional solutions rely on sparse initial interactions or demographic approximations, approaches that often fail to capture authentic user interests and preferences~\cite{kelly2009methods, bennett2012modeling}. Through the monitoring of cognitive responses during early system interactions, recommender systems could construct more accurate initial user models~\cite{eugster2016natural, davis2021collaborative}. This capability may have particular promise in domains where users possess limited domain expertise or struggle with preference articulation~\cite{belkin1982ask, marchionini2006exploratory}, potentially enabling more rapid convergence on effective personalisation strategies~\cite{jacucci2019integrating}.

Furthermore, real-time preference detection and adaptation are potential areas that may benefit from the integration of BMI systems ~\cite{jacucci2019integrating, eugster2016natural}. While contemporary systems have made significant advances through implicit feedback mechanisms~\cite{agichtein2006improving, joachims2017accurately}, they remain constrained by their reliance on observable behaviours, creating an inherent delay between preference formation and system adaptation~\cite{white2013enhancing}. BMI technologies offer unprecedented opportunities to bridge this gap through direct brain measurement of user responses~\cite{gwizdka2017temporal, allegretti2015relevance}. Prior art has demonstrated that brain signals can effectively differentiate between levels of user engagement, interest, and satisfaction~\cite{moshfeghi2013cognition, pinkosova2023moderating}, often preceding observable behavioural indicators by several hundred milliseconds~\cite{eugster2014predicting}. These capabilities suggest promising directions for developing more sophisticated and responsive recommendation algorithms that can adapt in real-time to users' cognitive states~\cite{davis2021collaborative, jacucci2019integrating}. Neurophysiological signal integration into recommender systems presents novel evaluation challenges that extend beyond traditional accuracy metrics~\cite{sanderson2010test, herlocker2004evaluating}. While established frameworks effectively measure ranking performance~\cite{jarvelin2002cumulated}, they cannot adequately capture the cognitive benefits of neurophysiological enhancement~\cite{kelly2009methods, gwizdka2019introduction}. Future research must develop evaluation methodologies that assess both system performance and cognitive alignment~\cite{eugster2016natural, jacucci2019integrating}, while addressing the technical complexities of real-time neurophysiological signal processing~\cite{lotte2018review, schirrmeister2017deep}.

\textbf{RD6:} \textit{Pursue BMI-enhanced recommender systems, with particular focus on addressing cold-start challenges, enabling rapid response to evolving user interests, and developing novel evaluation frameworks that capture both algorithmic performance and cognitive alignment.}

Neurophysiological signal integration into recommender systems presents novel evaluation challenges that extend beyond traditional accuracy metrics~\cite{sanderson2010test, herlocker2004evaluating}. While established frameworks effectively measure ranking performance~\cite{jarvelin2002cumulated}, they cannot adequately capture the cognitive benefits of neurophysiological enhancement~\cite{kelly2009methods, gwizdka2019introduction}. Future research must develop evaluation methodologies that assess both system performance and cognitive alignment~\cite{eugster2016natural, jacucci2019integrating}, while addressing the technical complexities of real-time neurophysiological signal processing~\cite{lotte2018review, schirrmeister2017deep}.

\subsection{Neurophysiological Enabled IR Systems}
\label{NeuroIR}
While Section \ref{Contextual} addressed how brain signals can enhance existing IR paradigms, this section explores transformative new interaction models that fundamentally reimagine information access. Rather than merely augmenting conventional search and recommendation systems, these approaches establish entirely new interaction paradigms where neurophysiological activity becomes the primary communication channel between users and information systems. Recent BMI developments have demonstrated the viability of direct brain pathways for system control~\cite{sitaram2017closed, ramadan2017brain}, suggesting opportunities for novel search interfaces that operate beyond traditional input mechanisms. Building on these advances, we propose a comprehensive framework for neurophysiological IR systems encompassing three complementary approaches: BMI-controlled systems enabling direct thought-based interaction~\cite{hochberg2012reach, blankertz2016berlin}, neuroadaptive systems that dynamically respond to users' cognitive states~\cite{jacucci2019integrating, eugster2016natural}, and neuroproactive systems that anticipate information needs before explicit expression~\cite{moshfeghi2019towards, mcguire2024prediction}. Together, these approaches represent not incremental improvements but a fundamental shift in human-information interaction paradigms~\cite{white2013enhancing, gwizdka2019introduction}.

\subsubsection{BMI-Controlled IR}
BMI-controlled IR systems represent a direct application of BMI capabilities to search interaction~\cite{wolpaw2002brain, ramadan2017brain}. Recent advances in brain decoding have demonstrated increasingly sophisticated capabilities in translating brain signals into semantic representations~\cite{mitchell2008predicting, huth2016natural, Yang2024MADMM, Yang2024NeuSpeechDN}, enabling more natural query formulation~\cite{ye2024query, mitchell2008predicting}. This advance is particularly significant as modern brain decoding techniques can capture nuanced semantic relationships~\cite{huth2016natural, Yang2024MADMM}, potentially enabling more precise query expressions than traditional keyword-based approaches. BMI-control shows particular promise for complex information-seeking tasks that exceed the capabilities of traditional interfaces~\cite{marchionini2006exploratory, li2008faceted}. However, implementing brain control for IR systems presents distinct technical challenges beyond basic interface manipulation~\cite{hochberg2012reach, lotte2018review}. Search interaction requires more sophisticated control mechanisms to support result navigation, dynamic filtering, and precise selection~\cite{blankertz2016berlin, sitaram2017closed}. These capabilities must maintain reliability across varying cognitive states while enabling efficient information exploration~\cite{gwizdka2017temporal, jacucci2019integrating}. The integration of real-time error correction and feedback mechanisms becomes particularly critical for maintaining search precision~\cite{schirrmeister2017deep, roy2019deep}. Research demonstrates that brain control mechanisms can enable more intuitive information exploration through continuous content scanning and dynamic filtering~\cite{ramadan2017brain, blankertz2016berlin}, yet achieving reliable and precise control for practical deployment presents significant technical hurdles~\cite{lotte2018review, schirrmeister2017deep}. Several fundamental challenges must be addressed to realize BMI-controlled IR systems: developing robust decoding methods optimized for search interactions~\cite{roy2019deep}, maintaining control precision across varying cognitive states~\cite{sitaram2017closed}, and establishing comprehensive evaluation frameworks~\cite{vabalas2020machine}. Critical to success is the integration of real-time error correction and feedback mechanisms that ensure system responsiveness while preserving user agency~\cite{jacucci2019integrating, eugster2016natural}. Addressing these challenges could transform information access, enabling more natural and efficient human-information interaction~\cite{marchionini2008human, white2013enhancing, 7878633}.

\textbf{RD7:} \textit{Design and evaluate novel direct brain-to-system interaction paradigms that enable hands-free search control, with particular emphasis on supporting users with motor impairments and optimizing the precision-versus-efficiency tradeoff in brain command interpretation.}

\subsubsection{Neuroadaptive IR}
Neuroadaptive IR systems introduce continuous cognitive state monitoring to enable dynamic optimization of search interfaces and retrieval strategies. These systems would aim to utilize real-time brain measurements to detect variations in cognitive load \cite{gwizdka2017temporal}, uncertainty \cite{moshfeghi2013understanding}, and engagement \cite{eugster2016natural}, creating opportunities for precise adaptation of the search experience. This capability may enable search systems to respond to users' cognitive states as they evolve throughout the information-seeking process \cite{ingwersen1996cognitive, kuhlthau2004seeking, michalkova2024understanding}. The adaptive mechanisms in these systems could operate across multiple dimensions of the search interaction. At the interface level, detection of increased cognitive load can trigger automatic adjustments to result in presentation, including simplification of layouts, modification of information density, and provision of additional contextual support \cite{gwizdka2010distribution, ji2024characterizing, kingphai2021eeg}. The adaptation extends to the underlying retrieval mechanisms, where the detection of user uncertainty may influence result diversity or trigger refinements to query understanding. A critical consideration in neuroadaptive systems is the significant variation in brain patterns and cognitive processing across users \cite{schirrmeister2017deep, lotte2018review}. Individual differences in brain signatures necessitate adaptive frameworks that can calibrate to each user's specific patterns while maintaining consistent performance \cite{vabalas2020machine, kostas2021bendr}. This personalization must address multiple system components, including signal interpretation, interface behaviour, and feedback mechanisms \cite{sitaram2017closed, bennett2012modeling, teevan2010potential}. Early research demonstrates that neuroadaptive systems can effectively interpret brain signals to infer relevant judgments \cite{eugster2016natural, eugster2014predicting} and detect cognitive load variations during search tasks \cite{gwizdka2010distribution, kingphai2021eeg}. The development of neuroadaptive IR systems requires careful consideration of the temporal dynamics of cognitive state changes and the corresponding system responses. The challenge lies not only in the accurate detection of cognitive states but also in determining appropriate adaptation strategies that enhance rather than disrupt the search process. This includes maintaining a balance between system responsiveness and stability, ensuring that adaptations support rather than interfere with users' natural information-seeking behaviours.

\textbf{RD8:} \textit{Create comprehensive frameworks for neuroadaptive IR systems that continuously monitor cognitive states, automatically adjust retrieval strategies, and dynamically modify interface elements while maintaining user agency and system explainability.}

\subsubsection{Neuroproactive IR}
Neuroproactive IR systems introduce anticipatory information delivery capabilities by detecting and responding to information needs before their explicit expression \cite{moshfeghi2019towards, mcguire2024prediction, michalkova2024understanding}. Building on advances in brain decoding of semantic information during query formulation \cite{mitchell2008predicting, huth2016natural, ye2024query} (discussed in section \ref{search-enhancement}), these systems may enable the identification of emerging information needs through continuous monitoring of brain signals, creating opportunities for pre-emptive information delivery. The approach offers particular value in time-sensitive contexts and complex tasks where maintaining cognitive flow directly impacts task performance \cite{knierim2025exploring, crescenzi2014too, gwizdka2010distribution}. 

The technical foundation of neuroproactive systems rests on two integrated processes: brain signature translation and contextually-aware information delivery. The first process involves continuous monitoring and interpretation of brain signals to identify patterns indicative of emerging information needs, leveraging the same semantic decoding techniques that enable enhanced query formulation \cite{mitchell2008predicting, huth2016natural, Yang2024MADMM}. The second process determines optimal timing and methods for information delivery based on the user's current cognitive state and task context \cite{white2009exploratory, hassan2014struggling}. Research has demonstrated the technical feasibility of this approach, with studies showing successful detection of information needs through brain signals approximately 500ms before conscious awareness \cite{eugster2014predicting, moshfeghi2019towards, michalkova2022information}. Beyond basic need detection, these systems show potential for assessing the urgency and complexity of information needs, enabling more sophisticated delivery strategies \cite{mcguire2024prediction, paisalnan2021towards}.

Implementation of neuroproactive IR systems presents distinct technical challenges that require careful consideration. Signal processing in real-world environments must address various sources of noise and interference while maintaining reliable detection of relevant brain patterns \cite{lotte2018review}. Individual variations in brain signatures necessitate robust adaptation mechanisms to ensure consistent performance across users \cite{schirrmeister2017deep}. The development of appropriate evaluation frameworks presents another significant challenge, as traditional IR metrics focused on post-query performance cannot adequately assess pre-emptive information delivery. New evaluation approaches must consider brain-state alignment, cognitive load reduction, and the effectiveness of anticipatory information provision \cite{sanderson2010test}. The ethical implications of neuroproactive systems require particular attention during development and deployment. The continuous monitoring of brain signals raises important questions about privacy and data protection, as these signals may contain sensitive personal information beyond immediate search-related patterns \cite{sitaram2017closed}.

\textbf{RD9:} \textit{Develop anticipatory IR systems that leverage brain signals to identify and satisfy emerging information needs.}

The success of neuroproactive IR systems ultimately depends on achieving an optimal balance between technical capability and human factors. While the challenges are significant, the potential benefits—including reduced cognitive load, improved task efficiency, and more intuitive information access for users with motor or cognitive impairments \cite{pasqualotto2015usability, lazarou2018eeg}—justify continued research and development in this direction.

\section{Challenges of BMI in IR}
\label{challenges}
The integration of BMI systems within IR presents transformative opportunities, yet realising this potential requires addressing several fundamental challenges. 
This section examines four core challenges that must be addressed to enable the practical implementation of BMI-enhanced IR systems.

{\bf Task-Specific Neuroimaging Challenges. }
The implementation of BMI-enhanced IR systems requires careful matching of neuroimaging capabilities to specific information-seeking tasks \cite{ramadan2017brain}. This presents unique challenges that go beyond basic technology selection, particularly when considering real-world deployment scenarios. A primary challenge lies in balancing measurement requirements across different search stages. Early-stage search processes, such as IN formation, require precise spatial localization to identify specific brain activation patterns \cite{moshfeghi2016understanding}. However, later-stage processes like relevance assessment demand high temporal resolution to capture rapid cognitive state changes \cite{eugster2014predicting}. This temporal-spatial tradeoff becomes particularly acute when attempting to support complete search sessions, where both types of measurement may be necessary. System deployment contexts create additional constraints. Laboratory-grade technologies that provide optimal measurements often prove impractical in realistic search environments due to cost, infrastructure requirements, and user acceptance factors \cite{mihajlovic2015wearable}. This creates a fundamental tension between measurement quality and practical usability that must be resolved for each specific application context. The challenge extends beyond single-technology solutions. While hybrid systems combining multiple neuroimaging modalities offer theoretical advantages, they introduce significant complexity in signal integration and real-time processing \cite{sitaram2017closed}. Questions of how to optimally combine and weight signals from different sources, particularly when they provide conflicting information, remain largely unresolved. These challenges are compounded by the diversity of search behaviours. Systems designed to support exploratory search may require sustained monitoring of cognitive load and engagement \cite{gwizdka2017temporal}, while those focusing on rapid information lookup need to prioritize detection of short-term state changes \cite{allegretti2015relevance}. Creating frameworks that can effectively match technology capabilities to these varied requirements while maintaining practical feasibility represents a critical research challenge.

\textbf{RD10:} \textit{Optimize neuroimaging technology selection for IR tasks through systematic evaluation.}

{\bf Real-time Brain Signal Analysis for IR. }
The dynamic nature of information seeking creates unique challenges for processing brain signals in BMI-enhanced IR systems \cite{lotte2018review}. These challenges extend beyond general signal processing concerns to address specific requirements of search interaction. Analysing brain signals during active search requires distinguishing between task-relevant cognitive processes and incidental brain activity. For example, when users examine search results, eye movements and reading-related brain patterns can mask the subtle signatures associated with relevance judgments \cite{gwizdka2017temporal}. Traditional artefact removal techniques often eliminate potentially valuable information about natural search behaviours, creating a tension between signal clean-up and preservation of meaningful search-related patterns. The temporal structure of search interactions presents particular processing challenges. Different search stages - from query formulation to result examination - occur at varying time scales and generate distinct brain patterns \cite{eugster2014predicting}. Processing approaches must adapt to these changing temporal dynamics while maintaining consistent performance. This becomes especially challenging when users rapidly switch between search sub-tasks, requiring systems to detect and process overlapping brain signatures. Current machine learning approaches for brain signal processing, while powerful, face significant limitations in IR contexts. Deep learning models have shown promise for decoding search-related brain patterns \cite{roy2019deep}, but their computational requirements often conflict with the need for real-time processing. Additionally, these models typically assume relatively stable brain signatures, whereas search behaviour can produce highly variable patterns depending on task complexity and user expertise \cite{moshfeghi2016understanding}. The need to maintain system responsiveness while ensuring signal quality creates particular challenges for BMI-enhanced IR.

\textbf{RD11:} \textit{Advance real-time brain signal processing techniques optimised for search interaction patterns.}

{\bf Evaluation Framework Development. }
The development and evaluation of BMI-enhanced IR systems are currently hampered by a lack of standardised, publicly available datasets that combine brain and search interaction data. Unlike traditional IR, where large-scale evaluation collections enable reproducible research, the collection of brain data during search tasks requires specialised equipment and careful experimental protocols. This has resulted in small, often incompatible datasets that limit the development of generalizable approaches. Furthermore, existing IR evaluation metrics may not adequately capture the cognitive benefits that BMI enhancement could provide. Traditional metrics focus on retrieval effectiveness and user satisfaction but may not reflect improvements in cognitive load reduction or the quality of proactive information delivery. This necessitates the development of new evaluation frameworks that can assess both traditional retrieval effectiveness and cognitive support capabilities \cite{gwizdka2010distribution}. The temporal nature of brain signals also presents unique challenges for evaluation. While traditional IR metrics often focus on discrete events like clicks or query submissions, brain signals provide continuous data streams that require new approaches to measurement and evaluation. Additionally, the relationship between brain states and search satisfaction may not be linear or easily quantifiable, requiring more sophisticated evaluation approaches.

\textbf{RD12:} \textit{Establish ethical frameworks and privacy protocols for BMI-enhanced information seeking.}

{\bf Privacy and Agency in BMI-Enhanced Search. }
The integration of brain measurements into IR systems introduces fundamental privacy and ethical challenges that go beyond traditional concerns about search data protection \cite{sitaram2017closed}. These challenges emerge from the unique nature of brain signals in information seeking contexts. Search behaviour inherently reveals users' knowledge gaps, interests, and decision-making processes. For example, brain signals could reveal uncertainty or confusion about search results before users consciously recognize these states \cite{moshfeghi2019towards}, raising questions about the boundaries between observable behaviour and private cognitive processes. The proactive capabilities enabled by brain measurements create particular ethical tensions in IR contexts. While early detection of information needs could improve search efficiency \cite{eugster2016natural}, it also raises concerns about system influence over users' natural information-seeking patterns. This becomes especially critical in scenarios where users are exploring new topics or forming opinions, as system interventions based on brain signals could potentially shape users' learning trajectories and belief formation. Search history has traditionally been considered sensitive personal data, but brain signals from search interactions add new dimensions to privacy considerations. These signals may reveal not just what information users seek, but their emotional responses, cognitive biases, and decision-making patterns during search \cite{gwizdka2017temporal}. This richness of personal data creates new risks for potential misuse, particularly in contexts where search behaviour might reveal sensitive personal characteristics or vulnerabilities. The long-term implications of BMI-enhanced search systems raise additional ethical concerns. Regular exposure to systems that can detect and respond to cognitive states may alter how users approach information seeking, potentially creating dependency on BMI-enhanced features or changing natural learning and discovery processes \cite{jacucci2019integrating}. These effects could be particularly significant for developing minds or in educational contexts.

\textbf{RD13:} \textit{Develop comprehensive evaluation methodologies for BMI-enhanced IR systems.}

\section{Summary}
\label{Summary}
This perspective paper has explored the transformative potential of BMIs in advancing IR beyond its current limitations. While traditional IR systems rely on explicit user input—queries, clicks, and interaction signals—to infer intent, they remain inherently reactive and unable to capture the deeper cognitive dimensions of information seeking. This constraint is particularly evident in high-complexity domains and among users who struggle to articulate their needs due to cognitive or physical barriers. By integrating BMI technologies, IR systems can transition from passive, interaction-driven models to proactive, cognition-aware retrieval. Neuroscientific research has demonstrated that search behaviours, including query formulation, relevance assessment, and cognitive load, are directly linked to measurable brain activity. Empirical findings suggest that BMI-enhanced IR models could refine search results dynamically, predict user frustration, and even anticipate emerging information needs before they are explicitly formulated. These capabilities have the potential to redefine IR, making search systems more intuitive, adaptive, and accessible.

BMI-powered IR presents a wealth of opportunities for future research and development. To fully realise the potential of cognition-aware search, foundational work must continue in understanding the neuroscience underlying IR, including the brain mechanisms involved in information seeking, decision-making, and relevance assessment. The opportunities identified through BMI integration directly address fundamental limitations in current IR methodologies. First, traditional IR systems struggle with accessibility, excluding users with motor impairments or those who find text-based interaction challenging. Second, current approaches cannot effectively handle ill-defined information needs, where users struggle to articulate their requirements through conventional queries. Third, existing IR methods rely on indirect behavioural signals that often misalign with users' true internal cognitive states, creating a persistent gap between system understanding and actual user intent. BMI integration offers promising solutions to these limitations through direct brain measurement, enabling more inclusive, intuitive, and cognitively-aligned information access. Equally important is the seamless integration of BMI with existing IR models, enabling systems to enhance their performance through real-time cognitive feedback. Furthermore, the development of entirely new IR paradigms designed to unlock the full potential of brain-driven search interactions could lead to fundamental advancements in human-information interaction.

This paper has outlined critical research directions that must be pursued to bridge these gaps, including the refinement of neurophysiological models for IR, the design of scalable and ethical brain-sensing technologies, and the implementation of cognition-aware retrieval frameworks. Despite these promising advancements, significant challenges remain. The integration of BMI into IR requires overcoming technical hurdles such as low signal-to-noise ratios in brain data, variability in brain responses across individuals, and the need for real-time processing. Additionally, ethical and privacy considerations surrounding brain data must be carefully addressed to ensure the responsible development and deployment of cognition-aware search systems. Future research must also establish robust interdisciplinary collaborations across IR, neuroscience, machine learning, and human-computer interaction to develop scalable, interpretable, and user-centric BMI-driven retrieval models. Ultimately, the integration of BMI into IR is not merely an enhancement of existing IR systems. It represents a fundamental transition in how humans interact with information. As BMI technology continues to evolve, IR systems will no longer be confined to explicit inputs but will instead become cognition-aware assistants capable of anticipating and responding to information needs at the level of thought itself. By setting forth this perspective, we provide a roadmap for the next generation of proactive IR systems, outlining the opportunities, challenges, and research directions needed to realise the full potential of cognition-aware search.

\balance
\bibliographystyle{ACM-Reference-Format}
\bibliography{ref}
\end{document}